# Vlasov simulation in multiple spatial dimensions

Harvey A. Rose[*†] and William Daughton[†]

A long-standing challenge encountered in modeling plasma dynamics is achieving practical Vlasov equation simulation in multiple spatial dimensions over large length and time scales. While direct multi-dimension Vlasov simulation methods using adaptive mesh methods [J. W. Banks et al., *Physics of Plasmas* 18, no. 5 (2011): 052102; B. I. Cohen et al., November 10, 2010, http://meetings.aps.org/link/BAPS.2010.DPP.NP9.142] have recently shown promising results, in this paper we present an alternative, the Vlasov Multi Dimensional (VMD) model, that is specifically designed to take advantage of solution properties in regimes when plasma waves are confined to a narrow cone, as may be the case for stimulated Raman scatter in large optic f# laser beams. Perpendicular grid spacing large compared to a Debye length is then possible without instability, enabling an order 10 decrease in required computational resources compared to standard particle in cell (PIC) methods in 2D, with another reduction of that order in 3D. Further advantage compared to PIC methods accrues in regimes where particle noise is an issue. VMD and PIC results in a 2D model of localized Langmuir waves are in qualitative agreement.

## I. INTRODUCTION

Particle trapping as manifested in equilibrium plasma configurations such as Bernstein-Greene-Kruskal modes[1], and its far from equilibrium counterpart, plasma wave breaking[2], are the essence of what distinguishes the kinetic from the fluid plasma regimes. In the collisionless regime, as described by the Vlasov equation[3], Particle-In-Cell (PIC) simulation methods[4] have long been a standard computational tool that manifests these phenomena. However, in some regimes, excess electric field fluctuations due to "particle noise" leads to unphysical solution properties. For example, recently it has been determined[5, 6] that many more PIC simulation particles (in 2D, typically the order of 512 particles per Debye length squared) are required than had been expected in stimulated Raman scatter (SRS) when trapped electron effects first become significant, as is the case near the SRS threshold[7]. This relatively large number of particles is required to suppress particle noise to the point where its unphysical contribution to the loss of electrons, trapped in the SRS daughter Langmuir wave, is a correction to physically dominant de-trapping mechanism. Since the latter is regime dependent, *e.g.*, electron-ion collision dominant regime versus a laser speckle side loss dominant regime versus a rapidly varying SRS regime, *a priori* estimates as to what constitutes a sufficient number of particles are not reliable and one must perform a convergence study[7] by adding more and more simulation particles. Other regimes, including instability onset, are also sensitive to noise levels. In particular, the Langmuir wave decay instability is easily disrupted by ion noise[8]. This contrasts dramatically with noise free Vlasov simulations whose multi-D application, though conceptually straightforward, may not be competitive with particle methods since 3D Vlasov solutions live in a six dimensional phase space, with hundreds of resolution points required in each of the three velocity directions, in each spatial cell. We note in passing, however, that simulation methods using adaptive mesh methods[9,10,11] have recently shown promising results in 2D. Such methods aside, PIC methods are already orders of magnitude faster than direct Vlasov simulation in 2D. Aside from the case of strongly magnetized plasma for which gyrokinetic ordering[12,13,14] applies, the authors are unaware of any other reduced multi-D Vlasov models whose solutions are also exact solutions of the original Vlasov equation. In this paper we introduce the Vlasov Multi Dimensional (VMD) model whose solutions are exact Vlasov solutions. The model introduced in Ref. 14 is the closest in spirit to our approach but it does not allow for self-focusing or side loss of trapped electrons both of which may be essential SRS saturation mechanisms, while the VMD model naturally allows both. The VMD model is appropriate in regimes for which waves propagate within a narrow cone of directions, as is the case for certain laser plasma instabilities (LPI) and Raman amplifier[15] short pulse generation. For such applications, 3D simulations can be a costly necessity. For example, NIF laser beams are polarization-smoothed[16], with intrinsically 3D geometry. Similarly, accurate modeling of the evolution of the SRS daughter Langmuir waves requires a 3D model because they are susceptible to self-focusing, a qualitatively different phenomenon[17] in 3D than in 2D.

---

[*]New Mexico Consortium, Los Alamos, New Mexico 87544, USA    hrose@newmexicoconsortium.org

[†] Los Alamos National Laboratory, Los Alamos, New Mexico 87545, USA

## II. THE VLASOV MULTIDIMENSIONAL MODEL

The primary VMD ansatz is that plasma waves propagate within a narrow cone of directions, whose axis will be called the "parallel" or "kinetic" or "wave" direction. For example, this direction may coincide with that of a large optic f#, $F$, random phase plate[18] laser beam, whose intensity fluctuations, "speckles", have a width (the perpendicular correlation length, $l_\perp$, of the laser beam in quiescent plasma) that scales as $F\lambda_0$, with $\lambda_0$ the laser wavelength. The narrow cone propagation ansatz requires that the plasma wave wavelength is small compared to $l_\perp$. The VMD model is motivated by the observation that for such waves, transverse electron motion removes electrons that might otherwise remain trapped in the troughs of nearly parallel propagating waves. A simple estimate of this de-trapping time scale or its inverse, the electron "side loss" rate[19], $\nu_{SL} = v_e/l_\perp$, with $v_e$ the electron thermal speed, may be compared with, *e.g.*, the rate at which collisional effects cause an electron to de-trap by diffusing in velocity an amount comparable to the velocity trapping width, $\sqrt{e\phi/m}$ with $\phi$ the Langmuir wave electrostatic potential amplitude. An improved $\nu_{SL}$ estimate would take into account the distribution of transverse speeds, $f(v_\perp)$, which in thermal plasma becomes ever more sharply peaked at $v \sim v_e$, $f(v_\perp) \sim v_\perp^{(D-2)} \exp(-v_\perp^2/2v_e^2)$, as the spatial dimension, $D$, increases. In 2D the distribution peaks at $v_\perp = 0$, while in 3D it peaks precisely at $v_e$, suggesting a model in which the continuum of transverse electron velocities is replaced by a discrete set whose magnitudes are peaked near $v_e$. Details of the distribution of particle speeds along the parallel direction, which are critical for accurate Landau damping and, nonlinearly, related trapping dynamics, must not be tampered with. The VMD model exactly retains parallel particle dynamics.

### *A. The VMD ansatz and dynamics*

Let *g* be a particular species' phase space distribution function. The VMD ansatz is:

$$g(\mathbf{x},\mathbf{v},t) = \sum_{i=1}^{N} f_i(\mathbf{x},v_\parallel,t)\delta[\mathbf{v}_\perp - \mathbf{u}_{i\perp}(\mathbf{x},t)] \quad (1)$$

A particle's velocity vector, $\mathbf{v}$, is represented by its transverse $\mathbf{v}_\perp$, and parallel projections $v_\parallel$, $\mathbf{v} = \mathbf{v}_\perp + v_\parallel \hat{\mathbf{e}}_\parallel$, with $\hat{\mathbf{e}}_\parallel$ the unit vector in the wave direction. All $f_i \geq 0$ since they represent particle densities. For technical reasons discussed below, the initial set of discrete transverse velocities are distinct, $\mathbf{u}_{i\perp}(\mathbf{x},t=0) \neq \mathbf{u}_{j\perp}(\mathbf{x},t=0)$, for all $\mathbf{x}$, if $i \neq j$. Consistency with basic plasma properties requires that the collection of transverse flow fields, $\{\mathbf{u}_{i\perp}\}$, have root mean square value of order the perpendicular thermal speed for each particle species. Each species' distribution function has such a representation but for simplicity only one species, "electrons", is explicitly shown. The number of transverse flow fields, $N$, and their initial and boundary values are mathematically arbitrary, but there must be at least two at finite perpendicular temperature in 2D and at least three in 3D.

If the perpendicular boundaries are open, escaping particles are replaced by, for example, a thermal distribution of electron velocities in the parallel direction, and by the physically appropriate choice of background flow in the perpendicular directions. In 3D, six flow components initially arranged uniformly in angle to form a hexagon, may be sufficient for isotropic wave propagation in the limit that the wave propagation direction makes a small angle with respect to the parallel direction, similar to the case of the 2D lattice gas[20] in which a hexagonal velocity distribution is necessary for isotropic hydrodynamics. Electric, $\mathbf{E}$, and magnetic, $\mathbf{B}$, fields are determined in standard fashion by charge and current densities, the sources in Maxwell's equations. For general initial $g$, the Vlasov equation, with units such that electron mass and charge for normalized to unity, determines $g$'s evolution by

$$\left\{\frac{\partial}{\partial t} + \mathbf{v}\cdot\nabla + [\mathbf{E} + (\mathbf{v}/c \times \mathbf{B})]\cdot\frac{\partial}{\partial \mathbf{v}}\right\}g = 0. \quad (2)$$

For the class of initial conditions given by Eq. (1), it determines the evolution of $\{f_i, \mathbf{u}_{i\perp}\}$ as follows: substitute Eq. (1) into Eq. (2) and integrate $\mathbf{v}_\perp$ about a neighborhood containing a particular $\mathbf{u}_{i\perp}$ to obtain

$$\left\{\frac{\partial}{\partial t}+v_\parallel \frac{\partial}{\partial x_\parallel}+\left[\mathbf{E}+(\mathbf{u}_{i\perp}/c\times\mathbf{B})\right]_\parallel \frac{\partial}{\partial v_\parallel}\right\}f_i+\nabla_\perp\cdot(\mathbf{u}_{i\perp}f_i)=0 \quad (3)$$

This step requires distinct flow fields. As a diagnostic, integrate Eq. (3) over $v_\parallel$ to obtain the standard continuity equation

$$\partial_t \rho_i + \nabla\cdot(\rho_i \mathbf{u}_i)=0. \quad (4)$$

with

$$\rho_i = \int f_i(\mathbf{x},v_\parallel,t)dv_\parallel, \quad (5)$$

and

$$\rho_i u_{i\parallel} = \int v_\parallel f_i(\mathbf{x},v_\parallel,t)dv_\parallel \doteq p_{i\parallel} \quad \mathbf{u}_i = \mathbf{u}_{i\perp}+u_{i\parallel}\hat{\mathbf{e}}_\parallel. \quad (6)$$

Note that the evolution of $f_i$ due to the $\partial/\partial x_\parallel$ term in the curly brackets of Eq. (3) results in density advection by $u_{i\parallel}$ in Eq. (4).

The current density is $\mathbf{j}_i = \rho_i \mathbf{u}_i$. Eq. (4) is not needed to evolve $\rho_i$ as it is already determined by Eqs. (3) and (5), but it is used to derive flow transport, Eq. (9). Integrate Eq. (2) over $v_\parallel$, multiply by $\mathbf{v}_\perp$ and again integrate over $\mathbf{v}_\perp$ in a $\mathbf{u}_{i\perp}$ neighborhood to obtain, with

$$\mathbf{p}_{i\perp} = \rho_i \mathbf{u}_{i\perp}, \quad (7)$$

$$\frac{\partial \mathbf{p}_{i\perp}}{\partial t}+\nabla\cdot(\mathbf{u}_i \mathbf{p}_{i\perp}) = \rho_i(\mathbf{E}+\mathbf{u}_i/c\times\mathbf{B})_\perp, \quad (8)$$

momentum transport in conservation form. It follows from Eqs. (4) and (8) that

$$(\partial/\partial t + \mathbf{u}_i \cdot \nabla)\mathbf{u}_{i\perp} = \mathbf{E}_\perp + (\mathbf{u}_i/c\times\mathbf{B})_\perp. \quad (9)$$

When *g* evolves according to the Vlasov equation, it will maintain the form given by Eq. (1), if Eq. (9) in fact has a solution. This would *not* be the case if perpendicular shocks formed. We will later estimate the conditions for avoidance of such shocks. An analogous fluid representation has[21] been used to model transport along stochastic field lines, including collisional effects. Eq. (9) may be viewed as the Eulerian description of single particle dynamics, accelerated in the perpendicular plane by electric and magnetic fields. Its velocity, $\mathbf{u}_{i\perp}$, is not convected by fluctuations in $v_\parallel$ because the former has no fluctuations as per the VMD ansatz, Eq. (1). This fact implies a quantitatively different linear dispersion relation than what usually follows from Vlasov dynamics, as discussed in section II.C.

*Equations* (3) *and* (9) *constitute the VMD model*, with Eq. (1) providing the connection to the standard phase space density. The model must be further constrained, however, to avoid unphysical two-stream instability and shocks, as discussed in the next paragraph. Its solutions are exact solutions to the Vlasov equation, with **E** and **B** fields determined by charge $\rho = \sum_i \rho_i$ and current $\mathbf{j} = \sum_i \mathbf{j}_i$ densities, plus Maxwell's equations, as usual. Recall that the summation over "$i$" is a sum over flow field components for a given species. An additional sum over species is required to obtain the total charge and current densities. Initial and boundary conditions for the $\{\mathbf{u}_{i\perp}\}$ are so far arbitrary, but we will show that this arbitrariness is largely resolved by invoking fundamental physical constraints such as isotropy and linear stability of fluctuations about thermal equilibrium. The initial flow distinctness constraint, while technically necessary to derive the VMD model from the Vlasov equation, is no longer required: solutions of the VMD model are solutions of the Vlasov equation.

If VMD dynamics leads to perpendicular shocks from smooth initial conditions, then the VMD model is not useful. Multiple valued flow fields could be introduced[22] but we choose not to follow this route. Perpendicular shocks essentially break the VMD ansatz, and the basic Vlasov equation with smooth distribution functions must be re-invoked. We now offer a heuristic argument as to when VMD dynamics does not lead to such shocks. Since self-advection is the key hydrodynamic shock mechanism, the parallel advection term is omitted as not increasing the tendency to shock. The Lorentz force is dropped for simplicity. Let there be one transverse direction, "$x$". One obtains for each flow index (suppressed), in lieu of Eqs. (4) and (8),

$$\partial_t \rho + \partial(\rho u)/\partial x = 0$$
$$\frac{\partial}{\partial t}(\rho u) + \frac{\partial}{\partial x}(\rho u^2) = \rho E_x \qquad (10)$$

$E$ depends on the total density. Dawson[23] has shown that this zero temperature plasma fluid model with sinusoidal initial conditions does not shock unless the initial relative density fluctuation is large compared to unity, a regime that is outside the scope of the VMD model. We offer this merely as a suggestion as to when VMD perpendicular shocks can be avoided. Solutions which vary rapidly along the parallel direction are compatible with the VMD *ansatz*, and since their dynamics is controlled by the Vlasov part of VMD (that contained in the curly brackets of Eq. (3)), such parallel "shocks" are ultimately smooth. Based upon our numerical simulations and stability analysis that follow, of greater concern than shocks is unphysical two-stream instability (TSI) caused by opposing flows in the VMD "thermal equilibrium" state. That state is characterized in the plasma rest frame by spatially uniform $\mathbf{u}_{i\perp}$, $\rho_i$ and $f_i$

$$f_i(x, v_\parallel) \propto f_0(v_\parallel/v_e, v_e) = \exp(-v_\parallel^2/2v_e^2)/\sqrt{2\pi}\, v_e, \tag{11}$$

with $\sum \rho_i \mathbf{u}_{i\perp} = 0$. If TSI is realized, it should be considered unphysical since the thermal equilibrium state must be stable. In the case of electron plasma with neutralizing static ion background, we will show in section II.D that TSI may be avoided if density fluctuations are limited to **k** such that $\theta_\mathbf{k} = \tan^{-1}(k_\perp/k_\parallel) < \theta_{max} \approx 0.6$, with $k_\parallel$ ($k_\perp$) the magnitude of the parallel (perpendicular) component of the Fourier transform wavevector, **k**. If initially there is not have much energy near the angular cutoff, $\theta_{max}$, but VMD evolution results in significant energy at $\theta_{max}$, then the underlying Vlasov model and VMD solutions are expected to have fundamental differences. For the examples discussed in this paper, the angular cutoff is self-consistently maintained for long times.

## *B. Simpler models and flow angular cutoff*

In the following discussion, coupling to the magnetic field, **B**, is ignored and the electrostatic regime assumed. The narrow wave cone ansatz suggests that

$$-\rho = \Delta\phi \approx \frac{\partial^2 \phi}{\partial x_\parallel^2} \tag{12}$$

is a useful approximation to Poisson's equation. Electron thermal units are used here and in the remainder of the paper: density is normalized to the background electron density, length to the electron Debye length and frequency to the electron plasma frequency. These units imply that the unit of speed is $v_e$, while the factor of $4\pi$ that multiplies the charge and current density in Gauss's law and Ampère's law respectively (in Gaussian units) is replaced by unity. If $\phi$ has a characteristic wavenumber, $k_\parallel$, with little energy in its harmonics, then the yet simpler harmonic model,

$$\phi \approx \rho / k_\parallel^2 \qquad (13)$$

may be employed. Since $\mathbf{E} = -\nabla\phi$, Eq. (8) now resembles coupled isothermal fluids with pressure proportional to $\rho^2$. They are coupled since the density, $\rho$, in Eq. (13) is the total density. Though these simplifications break the correspondence between VMD and Vlasov solutions, they provide insight into qualitative solution properties and provide a simpler test bed for numerical solution of the VMD, a nonstandard plasma dynamics model. In examples presented in section III, the full Poisson equation is used and Eq. (13) is verified *a posteriori*. With or without these simplifications, an angular cutoff applied to density fluctuations alone is not sufficient for basic physical fidelity of VMD solutions in the nonlinear regime. Consider the short system regime so that electric field fluctuations at harmonics of $k_\parallel$ are negligible. Then the angular constraint on density fluctuations becomes the simpler $k_\perp < k_{max} = k_\parallel \tan(\theta_{max})$, setting to zero $\mathbf{E}$'s Fourier mode components with $k_\perp > k_{max}$. If an initial condition is chosen for $\mathbf{u}_{i\perp}$ in Eq. (9), with Fourier support at wavenumbers $k > k_{max}$, then its evolution is unchecked by density compression and associated electric field which are limited to $k < k_{max}$, and unphysical shocks in general will develop. Thus a corresponding angular cutoff must be applied to the perpendicular flow field, $\mathbf{u}_{i\perp}$.

## C. Differences between linearized VMD and Vlasov solutions

8                                    Rose & Daughton                                    7/29/11

Typical initial conditions for the Vlasov equation are smooth, including that obtained from Eq. (1) if the delta function were softened. However, no matter how narrow, but finite, the modified delta function's width, such initially tiny diameter disks in phase space will spread and filament due to shear and, when coarse grained after long time evolution, will appear smooth. However, if $g$ is initially a sum of actual (singular) delta functions, then so it will remain. These two limits, narrowing the soft delta functions and evolving for long time, cannot be interchanged. One aspect of this difference appears in the comparison of linearized VMD versus Vlasov dynamics, discussed below. A related dichotomy has been noted[24] in the contrasting statistical dynamics of the Klimontovich versus Vlasov equations.

In 2D, let the parallel direction be "$z$", and the unique perpendicular direction "$x$". Formally initialize the Vlasov equation, Eq. (2), with the singular equilibrium,

$$g_0(x,z,v_x,v_z) = \frac{1}{2}\left[\delta(v_x - u) + \delta(v_x + u)\right] f_0(v_z). \tag{14}$$

Alternatively this equilibrium may be represented, non-singularly, in VMD notation, Eq. (1), with the index "$i$" taking on the two values, denoted by "+" and "-",

$$(f_\pm)_0 = f_0/2, \quad (u_{\pm\perp})_0 = \pm u \tag{15}$$

Electrostatic fluctuations about $g_0$, varying as $\exp i(\mathbf{k}\cdot\mathbf{x} - \omega t) = \exp i(k_x x + k_z z - \omega t)$, *and evolving as per the linearized Vlasov equation*, satisfy the standard form dispersion relation,

$$1 = \iint \frac{g_0}{(\mathbf{k}\cdot\mathbf{v} - \omega)^2} d\mathbf{v} \tag{16}$$

A uniform, static, charge neutralizing ion species is assumed. Substitute for $g_0$ as given by Eq. (14) to obtain

$$4k^2 \cos^2\theta = Z'(\zeta_-/\sqrt{2}) + Z'(\zeta_+/\sqrt{2}) \tag{17}$$

$$\zeta_\pm \cos\theta = v_\phi \pm u\sin\theta \tag{18}$$

$Z$ is the plasma dispersion function[25], $v_\phi$ is the phase velocity, $v_\phi = \omega/k$, and $\tan\theta = k_x/k_z$. As $\theta \to 0$ (parallel propagation), $u$ becomes irrelevant and the standard[25] electron plasma wave dispersion relation, $2k^2 = Z'(v_\phi/\sqrt{2})$, is recovered, all of whose roots are stable. As $\theta \to \pi/2$, the well-known[26] cold plasma two-stream dispersion relation, $2k^2 = 1/(v_\phi + u)^2 + 1/(v_\phi - u)^2$, is recovered, which has unstable modes if $ku < 1$.

On the other hand, Poisson's equation, and linearized VMD dynamics, Eqs. (3) and (9) imply,

$$f_\pm = \frac{1}{2}(\pm k_x u + k_z v_z - \omega)^{-1}\left[(\rho_+ + \rho_-)\frac{k_z}{k^2}\frac{\partial}{\partial v_z} - k_x u_{\pm\perp}\right] f_0(v_z) \qquad (19)$$

$$(\pm k_x u - \omega)u_{\pm\perp} = -\frac{k_x}{k^2}(\rho_+ + \rho_-) \qquad (20)$$

Eqs. (5), (19) and (20) yield the dispersion relation

$$k^2 = \Xi(v_\phi, u, \theta) \qquad (21)$$

$$4\Xi = \left[Z'(\zeta_-/\sqrt{2}) + Z'(\zeta_+/\sqrt{2})\right] - \tan^2\theta\left[\frac{1}{\zeta_-/\sqrt{2}}Z(\zeta_-/\sqrt{2}) + \frac{1}{\zeta_+/\sqrt{2}}Z(\zeta_+/\sqrt{2})\right] \qquad (22)$$

Eq. (17) coincides with (21) and (22) for $\theta = 0$ and as $\theta \to \pi/2$, but differ for general values of $\theta$. The difference may be traced to the fact that the transverse flow fields are not convected by fluctuations in $v_\parallel$. This implies a term

$\sim f_0(v_z)/(\pm k_x u + k_z v_z - \omega)(\pm u k_x - \omega)$ when $u_{\pm\perp}$ is evaluated from Eq. (20) and substituted into Eq. (19), resulting in the $Z(\zeta/\sqrt{2})/\zeta$ terms in Eq. (22). The term

$\sim (\partial f_0/\partial v_z)/(\pm k_x u + k_z v_z - \omega)$ in Eq. (19) leads to the more familiar $Z'(\zeta/\sqrt{2})$ terms. Since all roots of Eq. (22) are stable for $\theta = 0$, continuity implies that stability is maintained over a finite θ range. We will show explicitly that this is the case for the TSI and establish stability limits on the VMD cone angle.

### *D. Optimal choice of VMD thermal equilibrium parameters*

The VMD electrostatic dispersion relation, Eqs. (18), (21) and (22), follows from the particular choice of thermal equilibrium model, Eq. (15). If $u=1$, then the mean square velocity along any direction is unity. Any other value of $u$ implies temperature anisotropy and hence[27] instability to electromagnetic fluctuations. On the other hand, certain physical environments are not sensitive to the inclusion of electromagnetic fluctuations, and if such fluctuations are ignored, other criteria may be used to choose $u$, such as the condition that the correct Langmuir wave dispersion relation is isotropic for small $\theta$. If four transverse flow fields are allowed in 2D then it may be possible to satisfy both temperature isotropy and isotropic Langmuir wave propagation. Alternatively, the additional fields could be used to enforce 2$^{nd}$ order isotropy for another mode, such as the electron acoustic mode.[28][29] Since the computational burden scales linearly with the number of flow components, such gains in physical fidelity come at a significant increase of computational cost. In this paper, only the two-flow-field model is considered explicitly.

We now show how to choose the flow speed $u$ such that the Langmuir wave dispersion relation is isotropic to 2$^{nd}$ order in the propagation angle, $\theta$, and establish limits on $\theta$ to assure two-stream stability. The Langmuir wave phase velocity (we now drop its subscript, $\phi$), $v(k,\theta) = \omega(k,\theta)/k$, from which $k(v,\theta)$ may be obtained, is given by Eqs. (21) and (22), with implicit dependence on $u$. These equations, plus the requirement of small $\theta$ isotropy, $\partial^2 k/\partial \theta^2 = 0$, imply for phase velocities $v \gg 1$

$$u^2 = 5v^2/(3v^2 + 21). \tag{23}$$

The approach to this limit is slow, as seen in Figure 1. Note that the Langmuir wave branch (the descending branch in Fig. 2) is an insensitive function of $\theta$ and $u$. For v greater than about 2.6, there is little change in $k$ with changes in $u$ or $\theta$. For $k\lambda_D = 0.42$

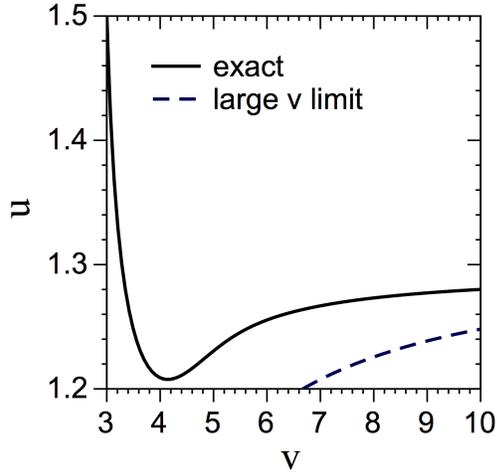 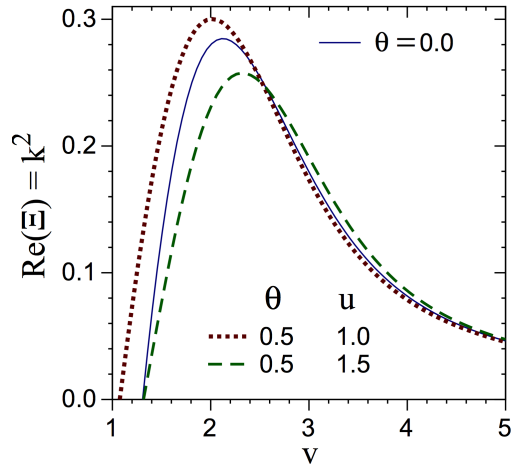

Fig. 1  Fig. 2

Fig. 1. 2$^{nd}$ order Langmuir wave isotropy requires that the initial transverse flow speed, $u$, vary with wave phase speed, v, solid curve. For large v, $u$ is given by Eq. (23), dashed curve. Both $u$ and v are in units of the electron thermal speed, $v_e$.

Fig. 2. The large v, v$\gtrsim$ 2.5,Langmuir wave, branch of the dispersion relation, Eqs. (21) and (22), is an insensitive function of the equilibrium flow speed, $u$, even for values of propagation angle, $\theta$, close to the two-stream instability boundary (see Fig. 4).

For $k\lambda_D = 0.42$ ( $v/v_e \approx 3$ ), variation with $u$ and $\theta$ of the Langmuir wave root of Eqs. (21) and (22), in the complex $\omega$ plane, is shown in Fig. 3. Recall that for $\theta = 0$, the parameter "$u$" drops out, recovering the standard dispersion relation.

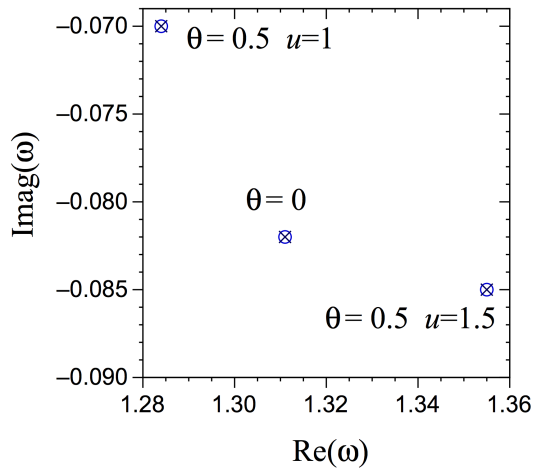 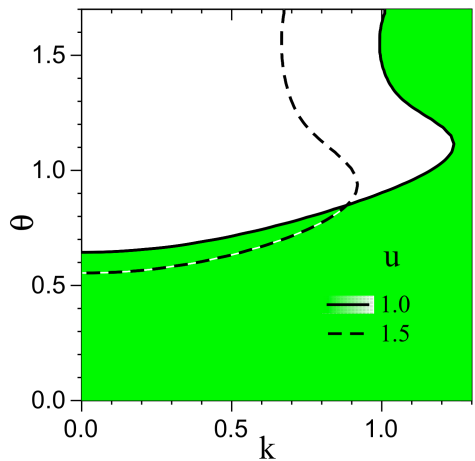

Fig. 3  Fig. 4

Fig. 3. Langmuir wave root (crossed circles) of the dispersion relation Eqs. (21) and (22) for $k\lambda_D = 0.42$ and various combinations of propagation angle, $\theta$, and equilibrium flow, $u$.

Fig. 4. The electrostatic dispersion relation, Eq. (22), for equilibrium flow speed $u/v_e = 1.0$ has two-stream stable modes for $k$ (in units of $1/\lambda_D$) to the right of the solid curve (green region), parameterized by the angle of propagation, $\theta$ (in radians). The dashed curve is the corresponding boundary for $u/v_e = 1.5$.

The generalized two-stream instability boundary depends on $u$. For $u = 1.0$, the solid curve in figure 4 is the continuation of the two-stream stability boundary from $\theta = \pi/2$ at which $k = 1/u = 1$. All points to the right of this curve, the green region in Fig. 4, are stable. The dashed curve in Fig. 4 is the stability boundary for $u = 1.5$. These stability boundaries are obtained by setting $v = 0$ in Eq. (22), which guarantees that the sum of the terms on the right hand side are real, and then finding $(k, \theta)$ pairs such that the real part vanishes.

## III. VMD and PIC SOLUTIONS COMPARED

Our choice of model for comparison of VMD and PIC simulation results adds a transversely localized traveling wave external potential, $\Phi_0$,

$$\Phi_0(x,z,t) = \phi_0 \cos[k_z(z - vt)] \times H_\phi(x). \quad (24)$$

In Eqs. (3) and (9), $\mathbf{E} \to \mathbf{E} - \nabla\Phi_0$, thus providing a source of Langmuir waves. The window function, $H_\phi$, localizes the Langmuir wave source symmetrically about $x = 0$ with length scale "$a$"

$$H_\phi(x) = \exp[-(x/a)^2].$$

In lieu of $a$, one may characterize its spatial extent by the full width at half maximum, $W = 2a\sqrt{\ln 2}$. If $\Phi_0$ were also localized in $z$, with length $L_z \gg W$, this could be

13              Rose & Daughton              7/29/11

interpreted as a simplified model of Langmuir waves driven by the beat ponderomotive force of laser and Raman scattered light in a large optic f# laser speckle[6]. While speckle stimulated Raman scatter is a physical motivation for this model, we choose the simplest geometry for our initial study: $L_z = 2\pi/k_z$ and periodic boundary conditions in z. Aside from dependence on the four parameters intrinsic to $\Phi_0$, there is time of evolution and initial and boundary conditions in x. Initial conditions are chosen as the constant density thermal equilibrium state with two initially equal but opposed flow components, $\pm u$. The boundary conditions for transverse advection are outgoing at the boundary $x = L_x/2$ for the positive flow component, with incoming $u_\perp$ and f pegged at their initial values at $x = -L_x/2$, and vice versa for the negative flow component. Poisson's equation is solved with doubly periodic boundary conditions as in the companion PIC simulations. The PIC code is described in Ref. 30. Outgoing boundary conditions were chosen to emulate an isolated speckle. It couples dissipatively with its environment since electrons escape and are replaced by those from the background thermal distribution.

The VMD numerical method is split step. First advance the Vlasov part of Eq. (3) (the terms in curly brackets) one time step, using the double Fourier transform method[31]. Next apply transverse advection to the $\{f_i\}$, and re-evaluate **E** with updated $\rho$. Lastly, update $\mathbf{u}_{i\perp}$ (Eq. (9)), with perpendicular and parallel advection. Fluctuations of $\mathbf{u}_{i\perp}$ are Fourier cutoff in angle as follows: at each z subtract the incident flow value, window the fluctuation, $\delta u(x) \to H_u(x)\,\delta u(x)$, so that it vanishes at the transverse boundaries and is hence a periodic function of x. Fourier transform the windowed fluctuation with respect to x and z, zero the modes with $\theta_\mathbf{k} > \theta_{max}$, inverse transform and finally add back the incident flow. More details are provided in the Appendix. Fig. 5 illustrates the two window functions, symmetric about $x = 0$, for the particular case $L_x = 450$. The flow window is a hyper-Gaussian to minimize its effects where $\delta u$ is near its maximum value, near $x = 0$:

$$H_u(x) = \exp\left[-14 \times 2^6 (x/L_x)^6\right]$$

14                                        Rose & Daughton                                    7/29/11

The multiplier, 14, in the exponent was chosen so that $H_u \approx 10^{-6}$ at the boundaries, $x = \pm L_x/2$, a dynamic range consistent with single precision arithmetic. The parameters

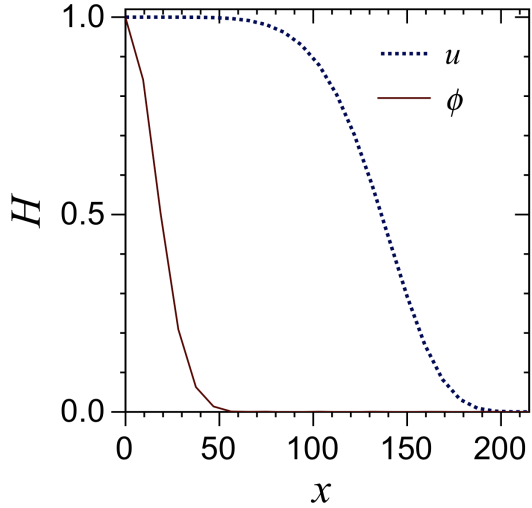
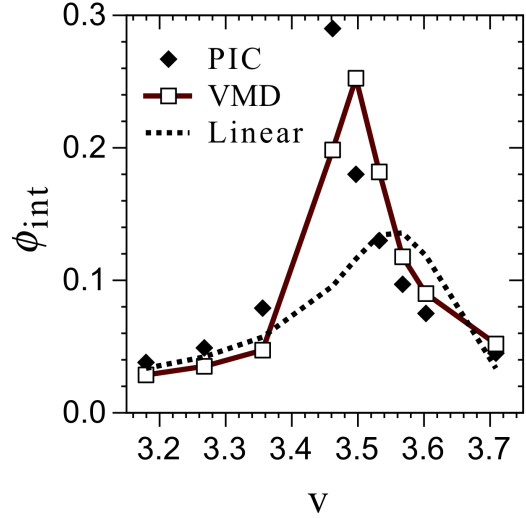

Fig. 5      Fig. 6

Fig. 5. Potential (solid) and flow (dotted) windows.

Fig. 6. Steady state Langmuir wave amplitudes compared: PIC (diamonds), VMD with $u$ varying with v to maintain isotropy (squares with solid connecting lines) and linear (dotted line) for $k_z = 0.34$, $\phi_0 = 0.01$ and $v_{SL} = 1/30$.

of this VMD/PIC results comparison are: $k_z = 0.34$, $\phi_0 = 0.01$, $\theta < 0.53$ and side loss rate $v_{SL} = \omega_{pe}/30$. This rate is fixed as the external potential's phase velocity varies. $W$ thus depends upon $u$, $W = u/v_{SL}$. Since the initial distribution of transverse speeds in the PIC simulations is thermal, it peaks at $v_\perp = 0$ and is more susceptible to trapping effects than the VMD model whose transverse speeds are at $u \approx v_e$. Therefore, one should not expect quantitative agreement with VMD solutions for given $v_{SL}$ if side loss is the dominant dissipative mechanism. Though for narrow enough speckle width, Langmuir wave diffractive losses will dominate[9], and side loss details are of lesser importance, this particular regime chosen for VMD-PIC comparison is intermediate, with neither of these loss mechanisms ignorable. Until the Langmuir wave amplitude, $\phi$, is large enough so that the electron bounce frequency, $\omega_{bounce}/\omega_{pe} = k\lambda_D \sqrt{e\phi/T_e} \approx \delta\rho/\rho$, exceeds $v_{SL}$,

Landau damping, $v_{Landau}$, is not significantly reduced by trapping[32], and Landau damping may be the dominant loss mechanism. Note that for the mid range values (v ≈ 3.5) of phase velocities shown in Fig. 6, $v_{Landau} \approx 0.035$.

In Fig. 6, $u$ varies with v as illustrated in Fig. 1, to ensure 2$^{nd}$ order isotropy. *W* varies with $u$ in VMD simulations to maintain fixed $v_{SL} = u/W$. Little difference was found when $u = 1$, independent of v, except for a slightly (10%) larger maximum response near v=3.5. Fig. 6 compares VMD, PIC and linear theory steady state values of $\phi_{int}$, the amplitude of the harmonic part of the self-consistent potential on axis, $x = 0$,

$$\phi(0,z) = \phi_0 \cos(k_z z) + \phi_{int} \cos(k_z z + \psi) + \ldots \qquad (25)$$

in the wave frame. The omitted terms in Eq. (25) represent the contributions from harmonics of $k_z$. Though negligible (energy less than 0.1% of that in the fundamental) they are retained in the simulations. The constant, $\psi$, allows for phase difference between the external and internal contributions to $\phi$. VMD solutions were followed in time long enough for the solution to become time independent in the wave frame. PIC solutions were evolved up to time of 500. Fig. 6 compares three evaluations of $\phi_{int}$: VMD, PIC and linear Vlasov theory

$$\phi_{int}/\phi_0 = \left| 1/\varepsilon(k_z, \omega = k_z v) - 1 \right|$$

Since the strength of the nonlinear Langmuir wave response and the phase velocity at maximum response (compare figures 4 and 5 in Ref. 19) depend on $v_{SL}$, VMD results for $v_{SL} = 1/30$, first shown in Fig. 6, are compared with $v_{SL} = 1/45$ (*W* increased by 50%) results, in Fig. 7. Aside from values of v near nonlinear resonance, v ≈ 3.5, there is no qualitative change in results. VMD and PIC evolution of $\phi_{int}$ are compared in Fig. 8 for $v = 3.53$, $v_{SL} = 1/30$ and two values of $u$. The PIC solution evolves more quickly because its relatively large number of slowly moving electrons experience more trapping oscillations than, *e.g.*, electrons moving at the thermal speed, while traversing the Langmuir wave core and hence trapping effects equilibrate more rapidly.

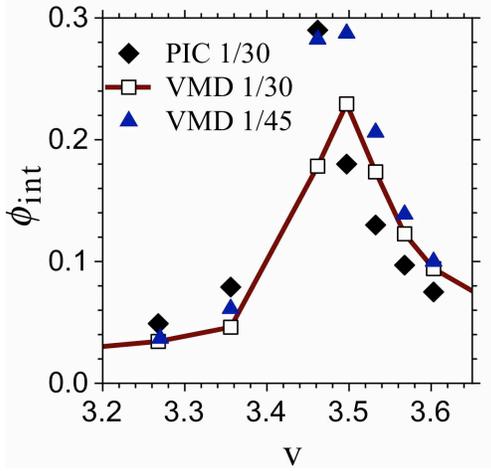
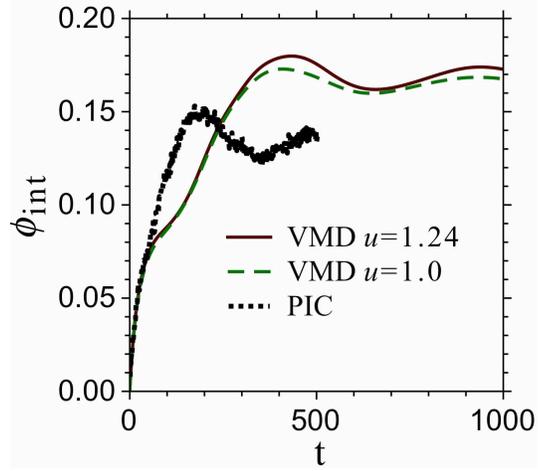

Fig. 7          Fig. 8

Fig. 7. Diamond and square symbol data are identical to those in Fig. 6, with additional VMD results (triangles) for a potential envelope whose width is 50% greater than the original. Symbols are labeled by their associated side-loss rate.

Fig. 8. VMD evolves more slowly than PIC in the approach to steady state Langmuir wave response for $v = 3.53$, $v_{SL} = 1/30$.

## IV. Langmuir Wave Bowing and Trapped Electron Self-Focusing

While the main purpose of this paper is introduction of the VMD model and preliminary comparison with PIC results, 2D solutions have a far richer structure than can be revealed solely by wave amplitudes. Here we present detailed VMD results for a wider system in which *a priori* estimates[9] suggest trapping effects stronger than diffraction, and hence wavefront bowing with positive curvature. The external potential's width, $W$, is increased to $133\lambda_D$, which is still small compared to a laser f8 optic speckle's diameter, $8\lambda_0 \approx 240\lambda_D$, for 1/10$^{th}$ critical density, 2.5keV plasma, with $\lambda_0$ the laser light wavelength. The external potential's parameters are $k_z = 0.34$, $\phi_0 = 0.01$ and $v = 3.50$. Initial state is VMD thermal equilibrium with $u = 1$. While the steady state value of $\phi_{int}$ is close to that of the much narrower system results shown in Fig. 7, solution geometries are qualitatively different. Steady state density fluctuation contours, Fig. 9, are concave up (in the direction of the external potential's phase velocity), while those of the narrower system (not shown) have opposite concavity. The former typifies a regime[9] where the Langmuir wave front is dominated by trapping rather than diffractive effects. Only the central portion of the entire simulation region, $|x| < 450$, is shown. The flow (potential) window has the same shape as in Fig. 5 but with twice (four times) the width. Deviations from uniform flow are small: magnitude less than 0.01 in the $x$ direction (compared with initial $u = 1$), and less than 0.13 in the $z$ direction (compared with the potential's phase velocity of 3.5). The $x$ component of the flow fluctuation, $\delta u_+$ about the positive flow component, $u_+ = 1 + \delta u_+$, shown in Fig. 10 appears smooth. This is quantified in Fig. 11, the graph of $|\delta\hat{u}(k,z)|^2$ ("^" denotes Fourier transform), averaged over $z$. We judge that the spectrum is compatible with the absence of shock formation, since the spectrum decreases rapidly as $k_{max} = k_z \tan(\theta_{max}) \approx 0.2$ is approached. This is consistent with a simple lower bound to the time of possible shock formation, $t_{shock}$, obtained by ignoring the electric field, thereby exaggerating the tendency to shock. One then estimates $t_{shock} \approx 1/|\partial u_+/\partial x| > 7{,}000$ from the data shown in Fig. 10, much larger than the time for advection of these fluctuations across the system by the background flow whose speed is unity.

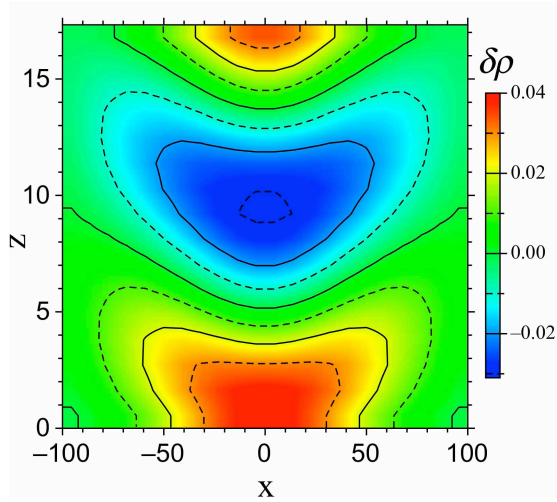 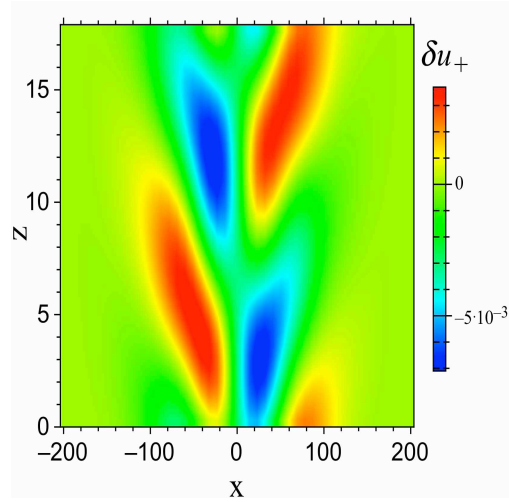

Fig. 9                                  Fig. 10

Fig. 9. Steady state density fluctuation for external potential width $133\lambda_D$ has wave fronts with positive concavity.

Fig. 10. Steady state positive flow fluctuation is smooth and small compared to electron thermal speed.

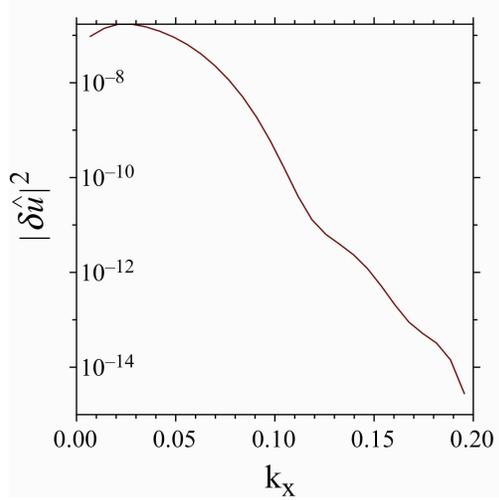 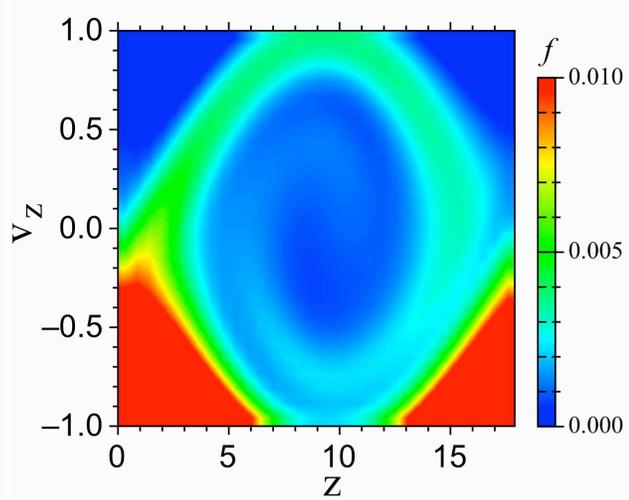

Fig. 11                                    Fig. 12

Fig. 11. Spectrum of $z$ averaged $x$ component of velocity fluctuation (arbitrary units) falls off quickly, suggesting that shock formation is not an issue.

Fig. 12. On axis phase space distribution function, $f_+(x=0)+f_-(x=0)$, in wave frame, shows only a few trapping oscillations.

The on axis (x=0) steady state phase space distribution function, $f = f_+ + f_-$, shown in Fig. 12, qualitatively deviates from a BGK mode in that contours of $f$ do not coincide with single particle energy contours, which are nearly symmetric under reflection about some line $z = cnst$ since the self-consistent potential is nearly sinusoidal. In a frame moving with the background flow, the number of oscillations, $N_{osc}$, of a trapped electron during the time it enters the system and travels to x=0 may be estimated as follows. From Fig. 9, $\omega_{\text{bounce}} \approx \sqrt{\delta\rho} \approx 0.2$, and the time during which it oscillates at this frequency before reaching the center is the density fluctuation half width (since $u = 1$). From Fig. 9, this distance is $\approx 50$, so that

$$N_{\text{osc}} \approx 0.2 \times 50/2\pi \approx 2$$.

Two trapping oscillations are not sufficient for a detailed correspondence with BGK mode properties: it is adequate for almost attaining the full trapped electron nonlinear frequency shift,[33] but several oscillations are required[34] to approach the trapping induced Landau damping reduction.

## V. Summary and Conclusions

The Vlasov Multi-Dimensional or "VMD" model, a fluid-Vlasov hybrid, has Vlasov dynamics along a preferred wave propagation direction and fluid dynamics in the perpendicular plane. Its solutions are exact solutions of the Vlasov equation. VMD achieves a computational advantage over standard particle in cell (PIC) or Vlasov methods by isolating intrinsic kinetic effects, such as electron trapping oscillations in a Langmuir wave trough, along the wave direction, and by allowing anisotropic spatial resolution with coarser grid in the perpendicular plane. The VMD model's practical utility is limited to regimes where the plasma waves are well collimated. Qualitative agreement between two-dimensional (2D) VMD and PIC simulation results was presented for Langmuir waves (with charge neutralizing, immobile ions) driven by a transversely localized source. Discrepancies may be understood as a consequence of PIC's thermal distribution of transverse speeds while VMD has discrete transverse speeds of order the electron thermal velocity, $v_e$, causing the latter to manifest weaker trapping effects in 2D. We expect this difference qualitatively reduced in 3D where the most likely transverse speed is $v_e$ for both models. Since the VMD model does not have the noise issues associated with PIC simulations, it has an additional computational advantage in simulating instability thresholds. The trapped electron filamentation instability, for example, may be studied with parameters similar to those used in our numerical study by initializing a uniform BGK mode and watching for the growth of fluctuations. Inclusion of ion dynamics, which is formally straightforward, will allow the study of competing thresholds, such as the Langmuir wave decay instability versus the trapped electron filamentation instability. While the VMD model cannot replace the Vlasov equation and its PIC simulations, it may allow relatively rapid exploration of qualitative solution properties in the collimated wave regime. The VMD simulations reported here were performed on one core of an Intel Core 2 Duo, 2.26GHz processor. The execution time is roughly $2\times10^{-4}$ seconds per $\lambda_D^2/\omega_{pe}$ with $dx \approx 10\lambda_D$, $dz \approx \lambda_D$, $dv_z \approx 0.1v_e$ and $dt = 0.1/\omega_{pe}$. Further economies and accuracy of computation are expected when more modern[35] 1D Vlasov simulation methods are used.

## Acknowledgements

We thank R. Berger, B. Cohen and H. Hittinger for helpful discussions and a preprint of their work (Ref. 9). H. Rose thanks B. Wendroff for discussions on basic hydrodynamic numerical methods and Natalia Vladimirova for discussions about 3D VMD coding. H. Rose is supported by the New Mexico Consortium and National Science Foundation award number 1004110.

## Appendix on numerical methods

Only electrostatics is considered here (set $\mathbf{B} = 0$). We use a variant of standard[36], split step, Fourier transform methods to evolve the Vlasov part of the VMD model, Eq. (3). Given $\mathbf{u}_{i\perp}$ at time $t$, advance $f_i$ sequentially first with the $v_\| \partial f_i/\partial x_\|$ term and then the $\nabla_\perp \cdot (\mathbf{u}_{i\perp} f_i)$ term. Update the density and evaluate $\mathbf{E}$ with angular cutoff. Next advance $f_i$ with the $E_\| \partial f_i/\partial v_\|$. Then evaluate $u_{i\|}$ and advance $\mathbf{u}_{i\perp}$ with Eq. (9), and apply the angular cutoff to $\mathbf{u}_{i\perp}$. Periodic boundary conditions are used in the parallel ($z$), and the phase space velocity, $v_z$, directions. The wave frame, $v-10 < v_z < v+10$, with $v_z$ the external potential's phase velocity, is used. This choice ensured a negligible magnitude for the distribution function at the velocity boundaries, consistent with periodic boundary conditions. For simulation results shown in Figs. 6, 7 and 8, the number of grid points (grid spacing) in the $x$, $z$ and $v_z$ directions are 48, 16 and 256 ($dx = 9.4 = dx_\perp/\lambda_D$, $dz = 1.2 = dx_{\parallel}/\lambda_D$ and $dv_z = 0.078 = dv_{\parallel}/v_e$) respectively, with time step $\omega_{pe} dt = 0.1$.
For Figs. 9, 10 and 11, the simulation width is doubled and correspondingly the number of points in the *x* direction increases to 96, while for Fig. 12 the number of points in the *z* direction was also increased to 32 and the time step decreased to 0.05 to maintain stability. Eq. (3) was modified by the addition of hyper "viscosity" terms,

$$\frac{\partial f}{\partial t} = \ldots - \left( D_v \frac{\partial^4}{\partial v_{\parallel}^4} + D_x \frac{\partial^4}{\partial x^4} \right) f.$$

*A priori*, $D_v$ is chosen such that for the largest value of the phase space velocity transform variable, $p = \pi/dv_{\parallel}$, the electric field shear is balanced, $Ep = Dp^4$, for the maximum value of *E* encountered during evolution. For the examples considered, $D = 10^{-6}$ proved adequate. For narrow enough systems, such as in the first example discussed in the body of the paper, the side loss rate, $v_{SL}$, is large enough to smooth the free streaming resonance, $k_z dv_z < v_{SL}$. When the external potential's width is increased by a factor of four, this inequality is no longer satisfied and finite $D_x$ is required with *a priori* estimate $k_z dv_z = D_x k_z^4$. Further adjustments of these hyper-viscosity coefficients may be required to ensure that their dissipation of *f* converges at the largest available value of their corresponding transform variables, but not so large as to render the highest modes dynamically superfluous. After each nonlinear contribution to the evolution, for example, $E \partial f/\partial v_{\parallel}$, *f* is de-aliased with the well-known 2/3 rule. Standard upwind finite differencing[37] is used for hydrodynamic advection. More sophisticated 1D Vlasov solutions methods are well known[35, 38, 39]. To avoid unphysical two-stream-instability, one must zero contributions to the electric field unless $\theta = \tan^{-1}(k_{\perp}/k_{\parallel})$ is below the stability boundary curve shown in Fig. 4, which for thermal background flow is approximately 0.65. In particular, electric field components that are uniform in the wave direction ($k_{\parallel} = 0$) are excluded. The transverse velocity fluctuations, evolved by Eq. (9), are correspondingly cutoff to avoid unphysical shock waves. For the particular solution geometry presented in sections III and IV, fluctuations at harmonics of $k_{\parallel} = k_z$ are strongly suppressed (recall that $k_z$ is the external potential's wavenumber) and the simpler conservative Fourier filter, $k_\perp/k_z < 0.6$ is used. Once it is determined that perpendicular shocks do not form, one may adopt a model in which the perpendicular advection contribution to Eq. (9) is linearized about its spatially uniform equilibrium value. Then it is no longer necessary to apply the angular cutoff to the flow fields as its application to the density fluctuation alone is sufficient to kill the unphysical two-stream instability. The flow window and Fourier analysis of the flow fields are now superfluous.

[18] Y. Kato et al., "Random Phasing of High-Power Lasers for Uniform Target Acceleration and Plasma-Instability Suppression," *Physical Review Letters* 53, no. 11 (1984): 1057.

[19] Harvey A. Rose and David A. Russell, "A self-consistent trapping model of driven electron plasma waves and limits on stimulated Raman scatter," *Physics of Plasmas* 8, no. 11 (November 2001): 4784-4799.

[20] U. Frisch, B. Hasslacher, and Y. Pomeau, "Lattice-Gas Automata for the Navier-Stokes Equation," *Physical Review Letters* 56, no. 14 (April 7, 1986): 1505.

[21] Harvey A. Rose, "Test-Particle Transport in Stochastic Magnetic Fields: A Fluid Representation," *Physical Review Letters* 48, no. 4 (January 25, 1982): 260.

[22] Jin-Gen Wang, G. L. Payne, and D. R. Nicholson, "Wave breaking in cold plasma," *Physics of Fluids B: Plasma Physics* 4, no. 6 (June 1992): 1432-1440.

[23] John M. Dawson, "Nonlinear Electron Oscillations in a Cold Plasma," *Physical Review* 113, no. 2 (January 15, 1959): 383.

[24] Harvey A. Rose, "Renormalized kinetic theory of nonequilibrium many-particle classical systems," *Journal of Statistical Physics* 20, no. 4 (April 1, 1979): 415-447.

[25] B. D. Fried et al., "LONGITUDINAL PLASMA OSCILLATIONS IN AN ELECTRIC FIELD," *J. Nucl. Energy, Part C* 1 (1960): 190.

[26] Dwight R. Nicholson, *Introduction to Plasma Theory* (John Wiley & Sons Inc, 1983).

[27] G. Kalman, C. Montes, and D. Quemada, "Anisotropic Temperature Plasma Instabilities," *Physics of Fluids* 11, no. 8 (1968): 1797-1808.

[28] Thomas Howard Stix, *The theory of plasma waves*, 1st ed. (McGraw-Hill, 1962).
25        Rose & Daughton        7/29/11

[29] D. S. Montgomery et al., "Observation of Stimulated Electron-Acoustic-Wave Scattering," *Physical Review Letters* 87, no. 15 (2001): 155001.

[30] William Daughton, Jack Scudder, and Homa Karimabadi, "Fully kinetic simulations of undriven magnetic reconnection with open boundary conditions," *Physics of Plasmas* 13, no. 7 (2006): 072101.

[31] Glenn Joyce, Georg Knorr, and Homer K. Meier, "Numerical integration methods of the Vlasov equation," *Journal of Computational Physics* 8, no. 1 (August 1971): 53-63.

[32] Harvey A. Rose, "Trapped particle bounds on stimulated scatter in the large k lambda[sub D] regime," *Physics of Plasmas* 10, no. 5 (May 2003): 1468-1482.

[33] G. J. Morales and T. M. O'Neil, "Nonlinear Frequency Shift of an Electron Plasma Wave," *Physical Review Letters* 28, no. 7 (February 14, 1972): 417-420.

[34] Thomas O'Neil, "Collisionless Damping of Nonlinear Plasma Oscillations," *Physics of Fluids* 8, no. 12 (December 1965): 2255-2262.

[35] A Ghizzo et al., "A Vlasov code for the numerical simulation of stimulated raman scattering," *Journal of Computational Physics* 90, no. 2 (October 1990): 431-457.

[36] Joyce, Knorr, and Meier, "Numerical integration methods of the Vlasov equation." *Journal of Computational Physics* 8, no. 1 (August 1971): 53-63.

[37] Charles Hirsch, "Numerical computation of internal and external flows. Vol. 2 - Computational methods for inviscid and viscous flows", 1990, http://adsabs.harvard.edu/abs/1990nyjw.book.....H.

[38] C.Z Cheng and Georg Knorr, "The integration of the vlasov equation in configuration space," *Journal of Computational Physics* 22, no. 3 (November 1976): 330-351.
26        Rose & Daughton        7/29/11

---

[39] F. Huot et al., "Instability of the time splitting scheme for the one-dimensional and relativistic Vlasov-Maxwell system," *Journal of Computational Physics* 185, no. 2 (March 1, 2003): 512-531.